\documentclass[12pt]{article}
\usepackage{amsmath,amssymb}
\usepackage{hyperref}
\usepackage{graphicx}              

\numberwithin{equation}{section} 

\newcommand{\Tr}{{\rm{Tr}}}

\begin{document}


\thispagestyle{empty}

 \renewcommand{\thefootnote}{\fnsymbol{footnote}}
\begin{flushright}
 \begin{tabular}{l}
 {\tt arXiv:0912.4664[hep-th]}\\
 \end{tabular}
\end{flushright}

 \vfill
 \begin{center}
 {\bfseries \Large Chern-Simons theory on  $L(p,q)$  lens spaces and Localization}
\vskip 1.9 truecm

\noindent{{\large
  Dongmin Gang \footnote{arima275(at)snu.ac.kr}}}
\bigskip
 \vskip .9 truecm
\centerline{\it Department of Physics and Astronomy,
Seoul National University,
Seoul 151-747, KOREA}
\vskip .4 truecm
\end{center}
 \vfill
\vskip 0.5 truecm

\begin{abstract}
Using localization technique, we calculate the partition function and the expectation value of Wilson loop operator in Chen-Simons theory on general lens spaces $L(p,q)$ (including $S^2 \times S^1$).  Our results are consistent with known results.
\end{abstract}

\vfill
\vskip 0.5 truecm

\setcounter{footnote}{0}
\renewcommand{\thefootnote}{\arabic{footnote}}

\newpage

\tableofcontents

\section{Introduction}
Some exact results on quantum field theories with fermionic symmetries can be derived using localization method. In \cite{Pestun:2007rz}, the method is  applied  to supersymmetric Wilson loop operators in $\mathcal{N}=4$ supersymmetric Yang-Mills theory and confirmed the conjecture by Erickson-Semenoff-Zarembo \cite{Erickson:2000af} and Drukker-Gross \cite{Drukker:2000rr}. Recently exact results for supersymmetric Chern-Simons theories on three sphere $S^3$ are derived using localization technique \cite{Kapustin:2009kz}. They obtain exact expression for partition function and expectation value of a supersymmetric Wilson loop in terms of matrix integral. As one application of their results, they reproduce the known results for partition function and Wilson loop expectation value in pure Chern-Simons theory on $S^3$.

In this paper, we apply the same localization method used in \cite{Pestun:2007rz,Kapustin:2009kz} to $\mathcal{N}=2$ supersymmetric Chern-Simons theories defined on lens spaces $L(p,q)$ ($p,q$ are coprime). We concentrate on the case with no matter, in which the theory reduces to pure Chern-Simons theory after integrating out auxiliary fields. Chern-Simons theory on lens spaces is studied in several contexts of physics, for example see \cite{Marino:2002fk,Aganagic:2002wv,Dolivet:2006ii,brini:2008,Griguolo:2006kp}.
Lens spaces $L(p,q)$ can be constructed by gluing two solid tori $D\times S^1$ together such that $(1,0)$ cycle of the first one is identified with $(q,p)$ cycle of the second. Here $(1,0)$ cycle denotes the contractible cycle in solid torus and $(0,1)$ denotes the other one. For example $L(0,1)=S^2 \times S^1$ and $L(1,0)=S^3$. This description for lens spaces is somewhat redundant, i.e.
\begin{align}
L(p,q)=L(p,q'), \quad \textrm{if  $q q'=\pm 1$ (mod $p$) or $q=\pm q'$ (mod $p$)}. \label{redundance}
\end{align}
Here the equality means a homeomorphism of manifolds. This kind of `surgery' can be generalized to construct
more general  three-manifolds. From this surgery description for 3-manifolds,
one can compute invariants (partition function and Wilson loop expectation value) of Chern-Simons theory on the spaces using its relation to two dimensional conformal field theory \cite{Witten:1988hf}.
It is found that the partition function of Chern-Simons theory on Seifert manifolds can be expressed as matrix integral \cite{Marino:2002fk}.

The organization of this paper is as follows: in section 2, we give brief review on relevant backgrounds. We explicitly write down the matrix integral formula for the partition function of Chern-Simons theory on $L(p,q)$ and Wilson loop expectation values on $L(p,-1)$. We review $\mathcal{N}=2$ Chern-Simons theory on general Riemannian 3-manifolds and the localization method used in \cite{Pestun:2007rz,Kapustin:2009kz}. In section 3 and 4, we calculate the partition function and the expectation value of Wilson loop operator in Chern-Simons theory on $L(p,q)$ using the localization and find exact matches with the known results. In section 5, we discuss further studies and difficulties of generalizing our methods to more general manifolds. Finally, appendices collect some useful results on spectrum of differential operators on three sphere and on monopole harmonics on two sphere.

\section{Reviews}
\subsection{Chern-Simons theory on lens spaces}
Consider a  Chern-Simons theory with gauge group $G$.
\begin{align}
S[A] = \frac{k}{4\pi}\int d^3 x  \Tr \big{(}\epsilon^{\mu\nu\rho}(A_\mu \partial_\nu A_\rho + \frac{2i}{3} A_\mu A_\nu A_\rho) \big{)}.
\end{align}
The theory can be defined on any three dimensional manifold $M$. We concentrate on the case $M=L(p,q)$.
We will consider $L(0,1)=S^2 \times S^1$ case and other lens space cases separately. For  $S^2 \times S^1$ case, the partition function and the expectation value for a Wilson loop is given by \cite{Witten:1988hf}
\begin{align}
Z^{S^2 \times S^1}(k) = 1, \quad \langle  W_{R} (C_{S^1}) \rangle = \delta_{R,0}. \label{know results 1}
\end{align}
Here $C_{S^1}$ denote a loop along the $S^1$ in the $S^2 \times S^1$. $R$ denotes a integrable representation in the current algebra for the gauge group $G$ at level $k$. For given $k$, there are only finite number of integrable representations. When $G=SU(2)$, representations with spin $s=0, \frac{1}2,\ldots ,\frac{k}2$ are integrable. For general $SU(N)$ case, integral representations at level $k$ are explained in section 3.  A Wilson loop operator along a closed curve $C$ is defined as
\begin{align}
W_R(C) = \Tr_R P \exp ( i \oint_C d \tau A_\mu \dot{x}^\mu). \label{Wilson_loop_pure_Chern-Simons}
\end{align}
Here $P$ denote the usual path-ordering operator. For other lens spaces $L(p,q)$, the partition function with gauge group $U(N)$ is given by \cite{hansen-2004-5} (ignoring overall factor which does not depends on $k$)
\begin{align}
&Z^{L(p,q)}(k) = \sum_m Z^{L(p,q)}(\hat{k},m) \quad \textrm{with }\hat{k}:= k+N,\nonumber
\\
&Z^{L(p,q)}(\hat{k},m) =  \frac{1}{(p \hat{k})^{N/2} } e^{\frac{i \pi}{\hat{k}} N(N^2-1) s(q,p) }  e^{i \frac{\pi \hat{k} q}p |m|^2} \sum_{\omega, \tilde{\omega}\in S_N} e^{\frac{2\pi i}{p \hat{k} } \omega(\rho)\cdot \rho - \frac{2\pi i}p \tilde{\omega}(m)\cdot (q \rho + \omega(\rho))}. \label{known results 2-1}
\end{align}
Here $\rho$ is the Weyl vector of $U(N)$, $\rho_i = \frac{N-2i +1}2 $,  and $\{m_i \}$ label flat gauge connections of the theory which will be explained in section 4.
$S_N$ denote the permutation group with $N$ elements and $|m|^2 := \sum_{i=1}^N m_i^2$.  Dedekind sum $s(p,q)$ is defined as
\begin{align}
s(q,p) = \frac{1}{4p} \sum_{j=1}^{p-1}\cot \frac{\pi j}p \cot \frac{\pi q j}p.
\end{align}
Following the procedure described in the section 3 of \cite{brini:2008}, we obtain
\begin{align}
&e^{i \frac{\pi \hat{k} q}p |m|^2} \sum_{\omega, \tilde{\omega}\in S_N} e^{\frac{2\pi i}{p \hat{k}} \omega(\rho)\cdot \rho - \frac{2\pi i}p \tilde{\omega}(m)\cdot (q \rho + \omega(\rho))} \nonumber
\\
&=(-1)^{N^2/2}(-ip\hat{k})^{N/2} e^{\frac{\pi i}{6p\hat{k}}N(N^2-1)} e^{\frac{\pi i \hat{k} (q+1)}p |m|^2} \nonumber
\\
& \times  \int d^N x e^{-i \hat{k} p \pi |x|^2 + 2 \pi \hat{k} m\cdot x} \prod_{\alpha>0}4 \sinh[\pi \alpha \cdot x] \sinh[\pi \alpha\cdot (x+  i \frac{ (q+1)}p m)], \nonumber
\\
&= (-1)^{N^2/2+N}(-ip\hat{k})^{N/2} e^{\frac{\pi i}{6p\hat{k}}N(N^2-1)} \nonumber
\\
&\times \int d^N x e^{-i \hat{k} p \pi |x|^2 + i \frac{\pi \hat{k} q}p |m|^2} \prod_{\alpha>0}4 \sinh[\pi \alpha \cdot (x+ \frac{i }p m)] \sinh[\pi \alpha\cdot (x-  i \frac{ q}p m)]
\end{align}
In the last line we change the variables $x_i\rightarrow -(x_i-  \frac{i}p m_i)$.
Here $\alpha>0$ denote the positive roots of $U(N)$ and diagonal matrices $x=\textrm{diagonal}(x_1 \ldots x_N)$ and $m=\textrm{diagonal}(m_1 \ldots m_N)$ are considered as the Cartan subalgebra of $U(N)$.
Dropping all the $k$ independent factors, the partition function becomes (we also ignore the framing dependent phase factor)
\begin{align}
&Z^{L(p,q)}(\hat{k},m)=\int d^N x e^{-i \hat{k}   \pi   (p |x|^2 - \frac{q}p |m|^2) } \prod_{\alpha>0} \sinh[\pi \alpha\cdot (x+\frac{i}p m) ] \sinh [\pi \alpha\cdot (x- i \frac{q}p m)]. \label{known results 2-2}
\end{align}
We consider a Wilson loop along a loop $C_{p,q}$ in $L(p,q)$ which corresponds to the generator (or its inverse) of the fundamental group $\pi_1 (L(p,q))=\mathbb{Z}_p$. See section 4.1 for its explicit form. For $q=-1$, the expectation value of the Wilson loop is given by following integration (see section 3.1 in \cite{Halmagyi:2007rw})
\begin{align}
\langle W_R (C_{p,q=-1})\rangle =  \frac{1}{Z^{L(p,-1)}} \sum_m  \int d^N x e^{-i \hat{k}   \pi   (p |x|^2 + \frac{1}p |m|^2) } \Tr_R \exp[2 \pi ( x+  \frac{i}p m)]  \prod_{\alpha>0}\sinh [\pi \alpha\cdot (x+ \frac{i}p m)]^2. \label{known results 2-3}
\end{align}
In section 3 and 4, we will rederive \eqref{know results 1},\eqref{known results 2-1},\eqref{known results 2-2} and \eqref{known results 2-3}  using localization.

\subsection{$\mathcal{N}=2$ Supersymmetric Chern-Simons theory on Riemannian three-manifolds}
In this paper we are considering  $\mathcal{N}=2$ supersymmetric Chern-Simons theory with gauge group $G$ defined on Riemannian three-manifold $M$.
To compare with the  pure Chern-Simons theory, we will concentrate on the case with no matter (chiral multiplet). The action for the theory is \cite{Kapustin:2009kz,Schwarz:2004yj} (we follow the convention in \cite{Kapustin:2009kz})
\begin{align}
S= \frac{k}{4\pi}\int_M d^3 x  \Tr \big{(}\epsilon^{\mu\nu\rho}(A_\mu \partial_\nu A_\rho + \frac{2i}{3} A_\mu A_\nu A_\rho) \big{)}  +\frac{k}{4\pi} \int_M d^3 x \sqrt{g} \Tr (- \lambda^\dagger \lambda + 2 D \sigma). \label{N=2 CS action}
\end{align}
Due to the gauge invariance, $k$ is integer-valued (we assume $k \geq 0$). If we integrate out the auxiliary fields ($\sigma,D, \lambda$), it reduces to pure Chern-Simons theory. The action is invariant under the following supersymmetric variation.
\begin{align}
&\delta A_\mu = \frac{i}{2} (\eta^\dagger \gamma_\mu \lambda - \lambda^\dagger \gamma_\mu \epsilon), \nonumber
\\
&\delta \sigma = - \frac{1}{2} (\eta^\dagger \lambda + \lambda^\dagger \epsilon), \nonumber
\\
&\delta D = \frac{i}{2} (\eta^\dagger \gamma^\mu (D_\mu \lambda) - (D_\mu \lambda^\dagger) \gamma^\mu \epsilon) - \frac{i}{2}(\eta^\dagger [\lambda, \sigma] - [\lambda^\dagger, \sigma]\epsilon)+ \frac{i}{6}(\nabla_\mu \eta^\dagger \gamma^\mu \lambda - \lambda^\dagger \gamma^\mu \nabla_\mu \epsilon), \nonumber
\\
&\delta \lambda = (-\frac{1}{2} \gamma^{\mu\nu}F_{\mu\nu}  - D + i \gamma^\mu D_\mu \sigma )\epsilon + \frac{2i}{3} \sigma \gamma^\mu \nabla_\mu \epsilon, \nonumber
\\
&\delta \lambda^\dagger = \eta^\dagger (\frac{1}{2} \gamma^{\mu\nu} F_{\mu\nu}- D - i \gamma^\mu D_\mu \sigma) - \frac{2i}{3} \sigma \nabla_\mu \eta^\dagger \gamma^\mu. \label{susy transformation}
\end{align}
Here $\epsilon$ and $\eta$ are arbitrary 2-component complex spinors.  We will consider the following supersymmetric Wilson loop along a closed curve $C$,
\begin{align}
W_R (C) =  \Tr_R  P \exp\big{(}\oint_C d \tau (i A_\mu \dot{x}^{\mu} + \sigma |\dot{x}|)\big{)}. \label{Supersymmetric Wilson}
\end{align}
Here $P$ denote the usual path-ordering operator. Note that after integrating out the auxiliary scalar fields, this operator becomes usual Wilson loop operator in pure Chern-Simons theory \eqref{Wilson_loop_pure_Chern-Simons}. The variation of this Wilson loop under the supersymmetry \eqref{susy transformation} is proportional to
\begin{align}
\delta W \propto - \eta^\dagger (\gamma_\mu \dot{x}^\mu + |\dot{x}|) \lambda + \lambda^\dagger (\gamma_\mu \dot{x}^\mu - |\dot{x}|)\epsilon.  \label{susy variatio for wilson loop}
\end{align}
For only certain loops $C$, the Wilson loop operator is invariant under the some supersymmetries.

\subsection{Localization method}
In section 3 and 4, we will use the localization method in \cite{Pestun:2007rz,Kapustin:2009kz}. Before that we will briefly summarize the method. Consider a quantum field theory with fermionic symmetry $\delta$.
\begin{align}
S[\Phi^i]=\int L[\Phi^i], \quad \delta S =0 , \quad \delta (\prod D \Phi)=0.
\end{align}
Here $\Phi^i$ denote the (fermionic or bosonic) fields in the theory and $ D\Phi$ is the path-integral measure.   Consider an operator $O[\Phi^i]$ which is invariant under the $\delta$, $\delta O[\Phi^i] =0$. Suppose that we deform the Lagrangian by adding a term $t \delta V [\Phi^i]$, which satisfies
\begin{align}
\int \delta^2 V =0 \textrm{ and } (\delta V)_{\textrm{bosonic}} \geq 0 .
\end{align}
Then one can argue that the partition function ($Z$) does not depend on $t$.
\begin{align}
\frac{d}{dt} \int D\Phi e^{i \int (L[\Phi^i]+t \delta V[\Phi^i])} = \int D \Phi \delta (i \int V[\Phi^i] e^{i\int L[\Phi^i]+ t \delta V[\Phi^i]}) =0.
\end{align}
Similar argument holds for the expectation value  $\langle O \rangle$ of the operator $O$. Thus we can take $t$ to be very large and the dominant contribution to the path integral will come from saddle points $\Phi_0$, which satisfy $\delta V (\Phi_0) =0$. We expand the deformed action ($L+t \delta V$) and operator $O$ around the saddle  points ($\Phi \rightarrow \Phi_0 + \frac{1}{\sqrt{t}}\Phi$).
\begin{align}
L[\Phi] + t \delta V[\Phi] \rightarrow L[\Phi_0]+\delta V_2 [\Phi; \Phi_0]+  o(\frac{1}{\sqrt{t}}), \quad O[\Phi] = O[\Phi_0]+o (\frac{1}{\sqrt{t}}). \label{quadratic expansion in general localization}
\end{align}
Here $\delta V_2$ denote the quadratic expansion of $\delta V$ in $\Phi$. If we take $t \rightarrow \infty$ limit, the path integral is simplified as
\begin{align}
&Z = \int d\Phi_0 e^{i S[\Phi_0]} (\int D\Phi e^{i \int \delta V_2[\Phi;\Phi_0]}) := \int d\Phi_0 e^{i S_0} Z_{1-loop}[\Phi_0], \nonumber
\\
&\langle O \rangle = \frac{1}{Z} \int d\Phi_0 e^{i S_0} Z_{1-loop} [\Phi_0]O[\Phi_0]. \label{partition function and operator in localization}
\end{align}
Thus the path integral is localized to the integration over saddle points $\Phi_0$.

\section{Localization :   $L(0,1)=S^2 \times S^1$ case }
In this section, we  consider a Chern-Simons theory  on $S^2 \times S^1$ with gauge group $G$. We assume that the group $G$  is path-connected. Then the set of inequivalent principal $G$-bundles over the manifold is in one-to-one correspondence with the element of $\pi_1(G)$. There are some subtlety when considering gauge connection in non-trivial $G$-bundle. A gauge connection(adjoint scalar) in the non-trivial bundle can't be represented by a Lie algebra valued one form(scalar) and the action \eqref{N=2 CS action} does not make sense. To avoid the problem we only consider a simply connected gauge group $G$. In particular, we choose $G=SU(N)$.
\subsection{Killing spinor and Supersymmetric Wilson loop}
We choose a metric of $S^2 \times S^1$ as
\begin{align}
ds^2_{S^2 \times S^1} = d\theta^2 + \sin^2 \theta d \phi^2 +  d \psi^2.
\end{align}
Here $\theta,\phi$ are the usual spherical coordinates and $\psi$ parameterizes the $S^1$,  $\psi \sim \psi+ 2 \pi$.
Some of Killing spinors in the orthonormal frame $\{d\theta, \sin \theta d \phi , d\psi \}$ are given by
\begin{align}
&\epsilon = e^{\psi/2}\left(
               \begin{array}{cc}
                 1 & 0 \\
                 0 & -i \\
               \end{array}
             \right)\cdot \exp(\frac{i}2 \theta \gamma_1)\cdot \exp(\frac{i}2 \phi \gamma_3) \cdot \epsilon_0
             , \nonumber
\\
&\epsilon^\dagger = e^{-\psi/2} \left(
               \begin{array}{cc}
                 1 & 0 \\
                 0 & i \\
               \end{array}
             \right)\cdot \exp(-\frac{i}2 \theta \gamma_1)\cdot \exp(-\frac{i}2 \phi \gamma_3)\cdot \epsilon_0^{*}, \label{killing spinor}
\end{align}
for constant $\epsilon_0$. These Killing spinors satisfy
\begin{align}
\nabla_\mu \epsilon = \frac{1}{2} \gamma_\mu \gamma_3 \epsilon, \quad \nabla_\mu \epsilon^\dagger = - \frac{1}2 \epsilon^\dagger \gamma_\mu \gamma_3.
\end{align}
We choose the gamma matrices as Pauli matrices $\gamma^i = \sigma^i$. We impose the following boundary condition for spinors
\begin{align}
&\lambda(\psi = \pi) = \exp(\pi ) \lambda (\psi= - \pi), \nonumber
\\
&\lambda^\dagger (\psi = \pi ) = \exp(- \pi ) \lambda^\dagger (\psi = -\pi ). \label{boundary for fermion}
\end{align}
Since fermion fields are auxiliary,  boundary condition for them has no physical meaning  and one can impose any consistent boundary condition on them. In Eulidean space fermion fields $\lambda$ and $\lambda^\dagger$ is treated as independent ones. We choose this boundary condition because the Killing spinors \eqref{killing spinor}, which play crucial role in localization,  satisfy this. Under the boundary condition  bosonic bilinears ($ \lambda^\dagger \gamma^{\mu_1}\ldots \gamma^{\mu_n}\lambda$) satisfy the periodic boundary condition, which is necessary for the theory to be invariant under the fermionic transformation.
\\
Consider the supersymmety transformation \eqref{susy transformation} generated by $\eta^\dagger =0$, and $\epsilon$=(Killing spinor).
\begin{align}
&\delta A_\mu = -\frac{i}2 \lambda^\dagger \gamma_\mu \epsilon, \quad \delta \sigma = - \frac{1}{2} \lambda^\dagger \epsilon , \quad \delta \lambda = - \frac{1}{2} \gamma^{\mu\nu}F_{\mu\nu}\epsilon - D\epsilon + i \gamma^\mu D_\mu \sigma \epsilon + i \sigma \gamma_3 \epsilon , \nonumber
\\
&\delta D = -\frac{i}2 D_\mu \lambda^\dagger \gamma^\mu \epsilon + \frac{i}2 [\lambda^\dagger, \sigma ]\epsilon - \frac{i}4 \lambda^\dagger \gamma_3 \epsilon. \label{susy variation}
\end{align}
We particularly choose $\epsilon$ as the Killing spinor \eqref{killing spinor} with $\epsilon_0 = (1,0)$. Then it satisfies
\begin{align}
\epsilon^\dagger \epsilon =1, \quad v^\mu \gamma_\mu \epsilon = \epsilon, \quad v^\mu v_\mu = 1, \quad   v^\mu \partial_\mu = \sin \theta \frac{\partial}{\partial \theta} + \cos \theta \frac{\partial}{\partial \psi},
\end{align}
where $v^\mu := \epsilon^\dagger \gamma^\mu \epsilon$. From \eqref{susy variatio for wilson loop} it can be shown that  the Wilson loop along the integral curve of $v^\mu$ preserves the supersymmetry. One simple integral curve of $v^\mu$ is $C_{S^1}$ which wraps the $S^1$ and located at $\theta=0$.

\subsection{Deformation and Saddle points}
We choose a fermionic variation $\delta$ generated by the Killing spinor  considered in the previous section. We deform the action by adding a term
\begin{align}
\delta V =  \delta \Tr \big{(}(\delta \lambda)^\dagger \lambda\big{)}.
\end{align}
Following the similar calculation in \cite{Kapustin:2009kz}, we obtain the following expression.
\begin{align}
\delta V = \Tr (\frac{1}{2} F_{\mu\nu}F^{\mu\nu} + D^2 + D_\mu \sigma D^\mu \sigma + \sigma^2 - \epsilon^{\mu\nu3}F_{\mu\nu} \sigma - i D_\mu \lambda^\dagger \gamma^\mu \lambda + i [\lambda^\dagger, \sigma ]\lambda - \frac{i}2 \lambda^\dagger \gamma_3 \lambda).
\end{align}
One can check that $\delta^2 V=0$ up to total divergences. Saddle points are determined by the equation $\delta \lambda =0$. That is
\begin{align}
&-\frac{1}2 \gamma^{\mu\nu}F_{\mu\nu} - D + i \gamma^\mu D_\mu \sigma + i \sigma \gamma_3 = 0 , \nonumber
\\
& \Rightarrow \frac{1}2 \epsilon_{\mu \alpha \beta}F^{\alpha\beta} = D_\mu \sigma + \delta_{3,\mu} \sigma , \quad D=0. \label{saddle point equation}
\end{align}
These equations are solved by
\begin{align}
&A= a(m)  + h_0 d\psi, \quad \sigma=\frac{m}2, \quad D=0, \label{monopole background}
\end{align}
for two commutating (traceless) hermitian matrices $m$ and $h_0$. Here $h_0$ represent holonomy along $S^1$, $A_\psi = h_0$ with $g(h_0):=\exp(2 \pi i h_0 )\in SU(N)$.
$a(m)$ represents a solution of Yang-Mills equation for $SU(N)$ gauge theory on $S^2$, which is a dirac monopole with charges $m\in su(N)$. Using the gauge symmetry, we choose both of $m$ and $h_0$ as diagonal matrices.
\begin{align}
&m= \textrm{diagonal}\{ m_1, \ldots, m_N \}, \quad h_0 =  \textrm{diagonal}\{ x_1, \ldots, x_N \}, \nonumber
\\
&\sum_i m_i =\sum_i x_i =0.
\end{align}
Explicit form of the monopole solution is given as
\begin{align}
a(m)&= -\frac{m}2 (\cos \theta -1)d \phi \quad \textrm{on the upper hemi-sphere},\nonumber
\\
&= -\frac{m}2 (\cos \theta +1)d \phi \quad \textrm{on the lower  hemi-sphere}.
\end{align}
Two local gauge potentials are glued together with the  transition function $g= e^{i m \phi }$ and all the adjoint fields are understood as sections of the associated bundle.\footnote{Since there is only one $SU(N)$ principal bundle over $S^2 \times S^1$, which is trivial bundle,  the bundle determined by $g=e^{ i m \phi}$ is actually equivalent to the trivial bundle with  $g=1$.
It means that the gauge potential of the dirac monopole can be written as (globally defined) $su(N)$  valued 1-form under a proper local gauge transformation.}
Due to the Dirac quantization, $m_i$ are integers and using the residual gauge symmetry, we may choose $m_1 \geq m_2 \dots \geq m_N$.
After taking into account the quantum shift of the Chern-Simons level $k\rightarrow \hat{k}:=k+N$,
the classical action \eqref{N=2 CS action} for these saddle points \eqref{monopole background} labeled by $(m,h_0)$ is\footnote{We have to be careful in calculating the classical action \eqref{N=2 CS action}, which is only well defined when a gauge field is a Lie algebra valued 1-form.
For more general connection $A$,  Chern-Simons part of the action can be defined as $\frac{k}{4\pi} \int_M \Tr F\wedge F$ where $\partial M = S^2 \times S^1$. }
\begin{align}
S_0 [m, h_0] = 2  \pi \hat{k} \Tr(m h_0).
\end{align}
The classical value of the supersymmetric Wilson loop along $C_{S^1}$  is
\begin{align}
W_R (C_{S^1})[m,h_0]= \Tr_R \exp(2 \pi i h_0 + \pi m ).
\end{align}
\subsection{1-loop determinant}
Expanding $\delta V$ around the saddle points as in \eqref{quadratic expansion in general localization}, we obtain (after dropping out overall trace and ignoring total divergence terms and integrating out an auxiliary field $D$)
\begin{align}
&\delta V_2  = (\bar{D}_1 A_2-\bar{D}_2 A_1)^2 + (\partial_3 A_i)^2 - \frac{1}{4}[A_i , m]^2 + \bar{D}_i \sigma \bar{D}_i \sigma + (\partial_3 \sigma)^2 + \sigma^2 -2 \sigma(\bar{D}_1 A_2-\bar{D}_2 A_1)  \nonumber
\\
&+ (\bar{D}_i \phi)^2 - \frac{1}{4}[\phi, m]^2  - i \partial_3 \sigma [\phi, m] + i \lambda^\dagger \gamma^i \bar{D}_i \lambda + i \lambda^\dagger \gamma^3 \partial_3 \lambda + i [\lambda^\dagger, \sigma]\lambda - \frac{i}2 \lambda^\dagger \gamma_3 \lambda  . \label{quadratic expansion}
\end{align}
Here $i,j=1,2$ indices represent the vielbein indices of the $S^2$ and $3$ represent the $S^1$. $\phi$ denote the 3rd component of gauge field, $\phi= A_\psi$. Effect of holonomy along the $S^1$, $h_0 d\psi$, is absorbed by redefining  adjoint fields, $\Phi \rightarrow e^{-i h_0 \psi}\Phi e^{i h_0 \psi}$.
\begin{align}
\partial_3 \Phi+ i [h_0,\Phi ] \rightarrow  e^{-i h_0 \psi}\partial_3 \Phi e^{i h_0 \psi}.
\end{align}
$\delta V_2$ in \eqref{quadratic expansion} is expressed in terms of newly defined fields and they satisfy following twisted boundary conditions.
\begin{align}
&\Phi(x^i, \psi=\pi)  = e^{ 2\pi i h_0}  \Phi(x^i, \psi=-\pi) e^{-2\pi i h_0}, \quad \textrm{for bosonic fields}, \nonumber
\\
&\Phi(x^i, \psi=\pi)  = e^{ 2\pi i h_0} e^{\pm\pi } \Phi(x^i, \psi=-\pi) e^{-2\pi i h_0}, \quad \textrm{for $\lambda$(plus sign) or $\lambda^\dagger$(minus)} . \label{twisted boundary conditions}
\end{align}
In \eqref{quadratic expansion}, we choose the  Coulomb (monopole) background gauge.
\begin{align}
\bar{D}_i A^i =0. \label{gauge fixing}
\end{align}
$\bar{D}_i$ represent  the covariant derivative on $S^2$ with the monopole background. The corresponding Faddeev-Popov determinant is $\det \bar{D}_i \bar{D}^i$.
As explained in the appendix B of \cite{Kim:2009wb}, the Coulomb gauge does not fix all the gauge redundance and the determinant for the residual gauge is given by
\begin{align}
\triangle(m,h_0) = \prod_{i<j; m_i = m_j}[ \sin (\pi (x_i - x_j))]^2,
\end{align}
up to overall factor. Note that this is nothing but the Haar measure on the broken symmetry group $H(m)\in SU(N)$ by the magnetic monopole with charge $m$, $[H(m),m]=0$.
We decompose all the adjoint fields $\Phi$ by  $\Phi = \Phi^0 + \Phi^\alpha X_\alpha$, where $\Phi_0$ represent the Cartan subalgebra part and $X_\alpha$ are representatives of the root space of $SU(N)$.
We choose the Cartan subalgebra $h$ as diagonal traceless hermitian matrices. $X_\alpha$ satisfies
\begin{align}
\Tr(X_\alpha X_\beta)= \delta_{\alpha,-\beta},\quad [h_0, X_\alpha]= \alpha(h_0)X_\alpha, \quad \textrm{$h_0 \in h$}.
\end{align}
Since the Cartan part $\Phi^0$ does not coupled to $(m,h_0)$ in \eqref{monopole background}, we will ignore it. For other fields $\Phi^\alpha$, the effective monopole charge is $q := \frac{1}{2} \alpha(m)$\footnote{$q$ in this section is nothing to do with $q$ in $L(p,q)$.} and the twisted boundary conditions \eqref{twisted boundary conditions} become
\begin{align}
&\Phi^\alpha(\psi=\pi)  =   e^{ 2\pi i \alpha (h_0)} \Phi^\alpha(\psi=-\pi), \quad \textrm{for bosonic fields}, \nonumber
\\
&\Phi^\alpha(\psi=\pi)  = e^{ 2\pi i \alpha (h_0)} e^{\pm\pi } \Phi^\alpha (\psi=-\pi) , \quad \textrm{for $\lambda$(plus sign) or $\lambda^\dagger$(minus)} . \label{twisted boundary conditions 2}
\end{align}
We expand bosonic fields in terms of monopole harmonics with charge $q=\frac{1}2 \alpha(m)$.
\begin{align}
&A^\alpha_i (x^i, \psi) = \sum_{n,j,m} a_{njm} \vec{V}_{qjm}(x^i) \frac{1}{\sqrt{2\pi}} e^{i(n+\alpha(h_0))\psi}, \nonumber
\\
&\sigma^\alpha(x^i , \psi) = \sum_{n,j,m} \sigma_{njm}Y_{qjm}(x^i)\frac{1}{\sqrt{2\pi}} e^{i(n+\alpha(h_0))\psi}, \nonumber
\\
&\phi^\alpha(x^i , \psi) = \sum_{n,j,m} \sigma_{njm}Y_{qjm}(x^i)\frac{1}{\sqrt{2\pi}} e^{i(n+\alpha(h_0))\psi}.
\end{align}
Here $\vec{V}_{qjm}$ and $Y_{qjm}$ are (divergenceless) vector and scalar monopole harmonics as explained in Appendix B. See \eqref{monopole harmonics 1},\eqref{monopole harmonics 2} for its properties. For $j\geq |q|+1$, the 1-loop determinant coming from these bosonic fields are
\begin{align}
& \prod_{j=|q|+1}^\infty \det \left(
       \begin{array}{ccc}
         \kappa_{qj}^2+q^2 + (n+ \alpha(h_0))^2 & \kappa_{qj} & 0 \\
         \kappa_{qj} & j(j+1)-q^2 + (n+ \alpha(h_0))^2+1 & (n+ \alpha(h_0)) q \\
         0 & (n+ \alpha(h_0)) q & j(j+1) \\
       \end{array}
     \right)^{2j+1} \nonumber
\\
&= \prod_{j=|q|+1}^\infty  ([j^2 + (n+\alpha(h_0))^2][(j+1)^2 + (n+\alpha(h_0))^2][j(j+1)-q^2])^{2j+1}.
\end{align}
For $j=|q|$ (in this case $\vec{V}_{qjm}$ is absent),
\begin{align}
\det \left(
       \begin{array}{cc}
         |q|+(n+\alpha(h_0))^2+1 & (n+\alpha(h_0))q\\
         (n+\alpha(h_0))q & |q|(|q|+1) \\
       \end{array}
     \right)^{2|q|+1} = \big{(}|q|[(|q|+1)^2+ (n+\alpha(h_0))^2]\big{)}^{2|q|+1}.
\end{align}
For $j=|q|-1$ (this is only possible when $|q| \geq 1$ and $Y_{qjm}$ is absent),
\begin{align}
(q^2+ (n+\alpha(h_0))^2)^{2|q|-1}.
\end{align}
Gathering all (including ghost contribution $\det\bar{D}_i \bar{D}^i  = \prod_{j=|q|}^{\infty} (j(j+1)-q^2)^{2j+1}$), the bosonic 1-loop determinant is
\begin{align}
&Z^{bos}_{1-loop}[m,h_0]= \prod_{\alpha,n} B[q=\frac{1}{2}\alpha(m),n,h_0]^{-\frac{1}{2}},
\end{align}
where $B[q,n,h_0]$ is defined by
\begin{align}
B[q,n,h_0]&:=[q^2 +(n+\alpha(h_0))^2]^{2|q|-1} \prod_{j=|q|+1}^{\infty}[j^2+(n+\alpha(h_0))^2]^{4j}, \quad \textrm{for }|q|\geq 1, \nonumber
\\
&:=\prod_{j=|q|+1}^{\infty}[j^2+(n+\alpha(h_0))^2]^{4j},\quad \textrm{for }|q| < 1.
\end{align}
We expand fermion fields $\lambda^\alpha$ in terms of monopole spinor with $q= \frac{1}2 \alpha(m)$.
\begin{align}
\lambda^\alpha = \sum_{n,j,m,\epsilon=\pm}   \lambda^\epsilon_{qjm} \Psi^\epsilon_{qjm}(x^i) \frac{1}{\sqrt{2\pi}} e^{i(n+\alpha(h_0))\psi} + \sum_{n,m} \lambda^0_{nm} \Psi^0_{q jm}(x^i) \frac{1}{\sqrt{2\pi}} e^{i(n+\alpha(h_0))\psi}.
\end{align}
$\lambda^\dagger$ is expanded in the same way. Here $\Psi^\epsilon_{qjm}$ are eigenspinors of Dirac operator in a monopole background whose properties are summarized in \eqref{monopole harmonics 3} and \eqref{monopole harmonics 4}.
For $j\geq |q|+ \frac{1}{2}$, one loop determinant from fermion interaction is
\begin{align}
\prod_{j=|q|+\frac{1}{2}}^\infty \det \left(
                               \begin{array}{cc}
                                 \mu_{jq}+ i q & - (n+\alpha(h_0))  \\
                                 -(n+\alpha(h_0))  & -\mu_{jq}+ i q  \\
                               \end{array}
                             \right) = \prod_{j=|q|+1}^\infty (-j^2 - (n+\alpha(h_0))^2)^{2j}
\end{align}
For $j=|q|- \frac{1}{2}$(this is only possible when $|q| \geq \frac{1}2{}$), the one-loop contribution is
\begin{align}
[iq + (n+\alpha(h_0))\textrm{sign}(q)]^{2|q|}.
\end{align}
Gathering all, the fermionic 1-loop determinant is
\begin{align}
Z^{fer}_{1-loop}[m,h_0]&= \prod_{\alpha,n}F[q=\frac{1}{2}\alpha(m),n,h_0]   , \nonumber
\end{align}
where $F[q,n,h_0]$ is given by
\begin{align}
F[q,n,h_0]&:= [q^2+(n+\alpha(h_0))^2]^{|q|} \prod_{j=|q|+1}^\infty [j^2+(n+\alpha(h_0))^2 ]^{2j}, \quad \textrm{for } |q|\geq \frac{1}{2} \nonumber
\\
&:=\prod_{j=|q|+1}^\infty[j^2+(n+\alpha(h_0))^2 ]^{2j}, \quad \textrm{for } q=0.
\end{align}
Thus 1-loop determinant is
\begin{align}
Z_{1-loop}[m,h_0] &= \prod_{\alpha,n} B[q=\frac{1}2 \alpha(m) , n, h_0]^{-\frac{1}{2}} F[q=\frac{1}2 \alpha(m) , n, h_0], \nonumber
\\
&= \prod_{\alpha,n;\alpha(m)\neq 0}[\frac{\alpha(m)^2}4 + (n+\alpha(h_0))^2]^{1/2}, \nonumber
\\
&=\prod_{\alpha >0;\alpha(m)\neq 0} [(\prod_{n=1}^\infty n^2)^2 (\frac{\cosh(\pi  \alpha(m))-\cos(2\pi \alpha(h_0))}{2\pi^2})].
\end{align}
Ignoring the overall infinity, we get
\begin{align}
Z_{1-loop}[m,h_0] = \prod_{\alpha>0; \alpha(m)\neq 0} [\cosh(\pi \alpha(m))- \cos(2 \pi \alpha(h_0))] .
\end{align}
The partition function and the Wilson loop expectation value is given by
\begin{align}
&Z^{S^2 \times S^1} = \sum_{m} \int_{T^{N-1}} dh_0 \triangle (m,h_0) e^{2 \pi i \hat{k} \Tr(m h_0)} Z_{1-loop}[m,h_0], \nonumber
\\
&\langle W_R (C_{S^1}) \rangle = \frac{1}{Z^{S^2\times S^1}}\sum_{m} \int_{T^{N-1}}   d h_0 \triangle(m,h_0) e^{2 \pi i \hat{k} \Tr (m h_0)} \Tr_R e^{2 \pi i h_0} Z_{1-loop}[m,h_0].
\end{align}
$T^{N-1}$ denotes the maximal torus of  $SU(N)$ and the integration over $T^{N-1}$ is defined as
\begin{align}
\int_{T^{N-1}} dh_0 f(h_0) :=  \int_0^1 dx_1 \ldots \int_0^1 dx_{N-1} f(x_1 , \ldots , x_{N-1}, x_N )|_{x_N = -\sum_{i=1}^{N-1}x_i}.
\end{align}
The $SU(N)$ character $\Tr_R e^{2 \pi i h_0}$ is given by
\begin{align}
\Tr_R e^{2 \pi i h_0} = \sum_{\lambda \in \Lambda_R } e^{2\pi i \lambda(h_0)}
\end{align}
where $\Lambda_R$ denotes the set of weights (including multiplicities) of representation $R$. Any element $\lambda$ in $h^*$ (dual vector space of Cartan subalgebra $h$) can be written as
\begin{align}
\lambda = \lambda_1 e_1 +\ldots \lambda_N e_N, \quad \sum_{i=1}^N \lambda_i =0.
\end{align}
Here $\{e_i \}_{i=1,\ldots N}$ is a complete set of $h^*$ defined by $e_i (h_0) = x_i$. For integrable representation $R$ of $SU(N)$ at level $k$, its highest weight $\lambda= \lambda_1 e_1 +\ldots + \lambda_N e_N$ ($\lambda_1 \geq \lambda_2 \ldots \geq \lambda_N $) satisfies the condition $\lambda_1 -\lambda_N \leq k$. For any weight $\lambda$ in $\Lambda_R$, it satisfy
\begin{align}
|\lambda_i | <k, \quad \lambda_i - \lambda_j  \in \mathbb{Z},\quad  \textrm{for any $i,j=1,2,\ldots, N$}.
\end{align}
We will show that for these integrable representation $R$,
\begin{align}
\int_{T^{N-1}}   d h_0 \triangle(m,h_0) e^{2 \pi i (k+N) \Tr (m h_0)} \Tr_R e^{2 \pi i h_0 + \pi m } Z_{1-loop}[m,h_0] = 0, \quad \textrm{if $m\neq 0$.} \label{lemma}
\end{align}
The integrand can be written as sum of following terms
\begin{align}
e^{2\pi i (c_1 x_1 + \ldots + c_N x_N)}, \quad c_1 + \ldots + c_N =0, \quad c_i - c_j \in \mathbb{Z}.
\end{align}
Note that if we integrate these terms over $T^{N-1}$, it vanishes unless $c_1 = \ldots =c_N =0$.
Therefore, if we expand the integrand as power series of $z := e^{2\pi i x_1}$ only the zeroth power of $z$ contribute the integration.
Let $P[f(z)]$ be the set of powers of $z$ in $f(z)$. For example, $P[3+2 e^{2\pi i x_1}+e^{-4 \pi i x_1}] =\{ 0,1,-2 \}$. One can see that
\begin{align}
&P[ e^{2 \pi i (k+N) \Tr (m h_0)}] = \{ m_1 (k+N) \}, \nonumber
\\
&|P[ \Tr_R e^{2 \pi i h_0 + \pi m }]| < k, \quad \textrm{if $R$ is a integrable representation at level $k$.}\nonumber
\\
&|P[ \triangle(m,h_0)Z_{1-loop}[m,h_0]]| \leq N-1.
\end{align}
Here $|S| <k$ means that $|s| < k$ for any element $s \in S$.  From these we can conclude that the integration vanishes unless $m_1 =0$. These argument can be straightly extended to other $m_i$ and we prove  \eqref{lemma}. Only the flat connections contribute the path integral. It is expected from the fact that only the flat connections are the stationary points of pure Chern-Simons action.  For $m=0$, $\triangle(m,h_0)$ is nothing but the Haar measure on $SU(N)$ and $Z_{1-loop}[m,h_0] =1$. Up to $k$ independent overall constant,
\begin{align}
&Z^{S^2\times S^1} = \int_{SU(N)} dg  = 1, \nonumber
\\
&\langle W_R (C_{S^1}) \rangle = \int_{SU(N)} dg \chi_R (g) = \delta_{R,0}.
\end{align}
$dg$ is the normalized Haar measure on $SU(N)$. These coincide with the known results \eqref{know results 1}.
\section{Localization :  $L(p,q) (p>0)$ case}
\subsection{Comparison with $S^3$ case}
The lens spaces $L(p,q)$ with $p>0$  can be considered as a $\mathbb{Z}_p$ quotient space of $S^3$. When we represent $S^3$ as  $|x|^2 +|y|^2=1$, the generator $e$ of the $\mathbb{Z}_p$ acts on $S^3$ as follows:
\begin{align}
e \cdot (x,y) = (\exp(\frac{2\pi i q}{p} )x, \exp(\frac{2\pi i }{p})y). \label{convention1}
\end{align}
We choose the metric of $L(p,q)$ as the same one with usual metric for unit $S^3$. Then the Killing spinor equations for $L(p,q)$ are the same with $S^3$ case at least locally. Using the same Killing spinor chosen in \cite{Kapustin:2009kz}\footnote{As in the $S^2 \times S^1$ case, we impose the boundary condition on spinor fileds such that the Killing spinor satisfy the condition. See the  last sentence in appendix A.}, which is constant spinor $\epsilon=(1,0)$ in the left-invariant frame,  we apply the same localization procedure to the $L(p,q)$ case. There are two main differences in $L(p,q)$ case  from $S^3$ case.

1. There are several discrete gauge inequivalent flat  connections on $L(p,q)$.

2. Spectrums of differential operators are different.
\\
For $S^3$ case, the supersymmetric Wilson loop along a great circle is invariant under the killing spinor. More concretely, we choose a great circle parameterized as follows.
\begin{align}
S(\theta) := \exp(i \sigma_3 \theta) \in SU(2), \quad 0 \leq \theta \leq 2\pi.
\end{align}
Here we use identification $S^3 = SU(2)$, see appendix A. For lens space cases, we consider a Wilson loop along the curve $C_{p,q}$ considered in the section 2. That is
\begin{align}
&C_{p,q} : \pi [ \{ S(\theta): 0\leq\theta\leq \frac{2\pi |q| }p )\} ].
\end{align}
Here $\pi$ is the projection map from $S^3$ to $L(p,q)$. The Wilson loop is invariant under the supersymmetry generated by the Killing spinor. In the next section, we follow each step for localization in $S^3$ case \cite{Kapustin:2009kz} and modify it suitably to the $L(p,q)$ case.

\subsection{Deformation and Saddle points }
First, we choose a deformation term $t \delta V$ as same one in \cite{Kapustin:2009kz}.
\begin{align}
\delta V = \Tr \big{(}\frac{1}{2}F^{\mu\nu}F_{\mu\nu}+D_\mu \sigma D^\mu \sigma +(D+\sigma)^2 +i \lambda^\dagger \gamma^\mu D_\mu \lambda + i [\lambda^\dagger, \sigma]\lambda - \frac{1}{2} \lambda^\dagger \lambda \big{)}. \label{delta V in S^3}
\end{align}
Saddle points satisfying $\delta V=0$ are given by
\begin{align}
F_{\mu\nu}=0 , \quad D_\mu \sigma =0, \quad \textrm{all other fields are  vanishing. } \label{fixed point equation}
\end{align}
For general 3 dimensional manifold $M$,  there are one-to-one correspondence between gauge inequivalent flat connections and homomorphism $H: \pi_1 (M )\rightarrow G$ up to conjugation. For $S^3$ the fundamental group is trivial and there's only one flat connection, trivial one. On the other hand, for lens space $L(p,q)$ the fundamental group is $\mathbb{Z}_p$ and there are several discrete flat connections. The $\mathbb{Z}_p$ can be identified with that in \eqref{convention1} in a natural way.  Suppose we choose the gauge group $G$ as $U(N)$.  Then flat connections on the lens space are labeled by diagonal matrices $m$ satisfying
\begin{align}
m=\textrm{diagonal}\{m_1, m_2, \ldots, m_N\}, \quad  p-1 \geq  m_i \geq 0, \quad m_1 \geq m_2 \geq \ldots \geq m_N .
\end{align}
The homomorphism $H : \mathbb{Z}_p \rightarrow U(N) $ for each $m$ is determined by
\begin{align}
H(e) = \exp(-\frac{2\pi i m}p ). \label{convention2}
\end{align}
For each choice of $m$, let the corresponding flat connection be $A_m$. If we write the flat connection as $A_m = - i g_{m}^{-1}dg_{m}$, then $g_m$ satisfies
\begin{align}
g_m \big{(}e\cdot (x,y) \big{)}  = \exp (-\frac{2 \pi i m }p ) g_m(x,y).
\end{align}
Holonomy of the flat connection along the loop $C_{p,q}$ is given by
\begin{align}
\Tr_R P \exp(i \oint_{C_{p,q}} A_m ) &= \Tr_R \exp(-\textrm{sign}(q) \frac{2\pi i m}p). \label{property of flat connection}
\end{align}
Here we use the fact that the loop $C_{p,q}$ corresponds to $e$ in $\pi_1 (L(p,q))=\mathbb{Z}_p$ for $q>0$ and to $e^{-1}$ for $q<0$. The first equation in  \eqref{fixed point equation} can be solved by choosing flat connection $A_m$ and the remaining equation for $\sigma$ can be solved by
\begin{align}
d \sigma +  [g_m^{-1}dg_m, \sigma] =0 \quad \Rightarrow \quad \sigma = g_m^{-1} \sigma_0 g_m.
\end{align}
Here $\sigma_0$ is a constant hermitian matrix.  Periodic boundary condition for $\sigma$ requires $\sigma_0$ to commute with $m$.
\begin{align}
\sigma \big{(}e\cdot (x,y)\big{)} = \sigma (x,y) \quad \Rightarrow \quad [\sigma_0 , m]=0.
\end{align}
Thus saddle points are labeled by two commuting matrices $(m,\sigma_0)$. Values of classical action \eqref{N=2 CS action} for the saddle points are (taking into account the quantum shift  $k\rightarrow \hat{k}:=k+N$)
\begin{align}
S_{0}[m, \sigma_0 ] = -\frac{\pi \hat{k} }p \Tr(\sigma_0^2) + \frac{\pi \hat{k}  q^*} p \Tr(m^2).
\end{align}
Here $0 \leq q^* \leq p-1$ is determined by the equation $qq^* = 1 (\textrm{mod } p)  $. The second part comes from the Chern-Simons action for flat connections (see the conjecture 5.6 in \cite{hansen-2004-5} and section 4.1 in \cite{Griguolo:2006kp})\footnote{Due to \eqref{redundance}, $q$ dependence of the Chern-Simons action for flat connections depends on  convention. In our convention \eqref{convention1},\eqref{convention2}, the classical action for the flat connections seems to be $\frac{\pi k q^*}p \Tr(m^2)$. In this choice the final partition function is consistent with \eqref{redundance}.}.
The classical value of  supersymmetric Wilson loop along the $C_{p,q}$ is
\begin{align}
&W_R (C_{p,q})[m,\sigma_0]= \Tr_R \exp[\textrm{sign}(q) \frac{2 \pi ( q \sigma_0 - i m  )}p].
\end{align}
\subsection{1-loop determinant}
The quadratic expansion $\delta V_2$ in \eqref{quadratic expansion in general localization} around the saddle points $(\sigma_0 , m) $ is (after integrating out $D$)
\begin{align}
\delta V_2 &= \Tr \big{(} \bar{D}_\mu A_\nu \bar{D}^{\mu}A^{\nu} +\bar{D}_\mu \sigma \bar{D}^{\mu}\sigma + [A_\mu , g_m^{-1}\sigma_0 g_m]^2  + i \lambda^\dagger \gamma^\mu \bar{D}_\mu \lambda + i [\lambda^\dagger, g^{-1}\sigma_0 g]\lambda  - \frac{1}2 \lambda^\dagger \lambda \big{)}.
\end{align}
$\bar{D}_\mu$ is the covariant derivative with the flat connection $A_m$. If we redefine all the adjoint fields $\Phi$ as follows,
\begin{align}
\Phi \rightarrow g_m^{-1} \Phi g_m ,
\end{align}
then $\delta V_2$ simplifies as
\begin{align}
\delta V_2 &= \Tr (\partial_\mu A_\nu \partial^{\mu}A^{\nu} +\partial_\mu \sigma \partial^{\mu}\sigma + [A_\mu , \sigma_0][A^\mu , \sigma_0 ] + i \lambda^\dagger \gamma^\mu \nabla_\mu \lambda + i [\lambda^\dagger, \sigma_0 ]\lambda - \frac{1}{2} \lambda^\dagger \lambda ) .\label{quadratic expansion 2}
\end{align}
In terms of newly defined fields the periodic boundary condition is modified as follows,
\begin{align}
\Phi(e \cdot (x,y)) = \exp ( -\frac{2\pi i m}p) \Phi (x,y)  \exp ( \frac{2\pi i m}p).
\end{align}
Here we observe the well known exchange between flat connections and twisted boundary conditions \cite{Hosotani:1988bm}. We expand adjoint fields by $\Phi=\Phi^0 +\Phi^\alpha X_\alpha$ as in section 3. Then, the boundary condition for each fields becomes
\begin{align}
&\Phi^0 ( e \cdot (x,y) ) = \Phi^0 (x,y), \nonumber
\\
&\Phi^\alpha  (e \cdot (x,y)) = \exp (-\frac{2 \pi i \alpha (m)}p ) \Phi^\alpha (x,y). \label{boundary condition}
\end{align}
We choose Lorentz gauge for newly defined gauge fields,  $\nabla_\mu A^\mu=0$. The 1-loop determinant for the quadratic expansion \eqref{quadratic expansion 2} is given by the same form in \cite{Kapustin:2009kz}, that is
\begin{align}
Z_{1-loop}[\sigma_0 , m] = \prod_\alpha \frac{\det (i \gamma^\mu \nabla_\mu + i \alpha (\sigma_0)- \frac{1}{2})}{\det(\nabla^2 + \alpha(\sigma_0 )^2 )^{1/2}}.
\end{align}
Here we ignore the fields which does not couple to $\sigma_0$ or $m$.  We also assume that $\sigma_0$ is in the Cartan subalgebra. The determinant are taken over divergenceless vector fields (or spinor fields) on $L(p,q)$ satisfying the boundary condition \eqref{boundary condition} for each $\alpha$. These spectra can be obtained from the spectra on $S^3$ with proper quotient. We assign a $U(1)_Q$ charge $Q$ on two complex number $(x,y)$ in \eqref{convention1} as follows
\begin{align}
Q(x)= q, \quad Q(y)=1.
\end{align}
Then a field $\Phi(x,y)$ with the $U(1)_Q$ charge $Q$ satisfies the following:
\begin{align}
\Phi \big{(}e\cdot (x,y)\big{)} = \exp(\frac{2\pi i Q} p ) \Phi(x,y).
\end{align}
Therefore, only the spectrum on $S^3$ with $Q + \alpha(m) \in p \mathbb{Z} $ for each $\alpha$ satisfies the boundary conditions \eqref{boundary condition}. Spectrums on $S^3$ and its $U(1)_Q$ charge are summarized in appendix A. Using them the 1-loop determinant is given by
\begin{align}
Z_{1-loop}[\sigma_0 , m] = \prod_\alpha \frac{ \prod_{l=0}^\infty [-l + i \alpha(\sigma_0)- 2]^{[(-\frac{l+2}2 , \frac{l}2) ,(-\frac{l}2, \frac{l}2)]_\alpha} \prod_{l=1}^\infty [l + i \alpha(\sigma_0)]^{[(-\frac{l}2, \frac{l-2}2),(-\frac{l}2 , \frac{l}2)]_\alpha} }{ \prod_{l=0}^{\infty}[(l+2)^2 + \alpha(\sigma_0)^2]^{\frac{1}2[(- \frac{l+2}2, \frac{l+2}2),(-\frac{l}2, \frac{l}2)]_\alpha+\frac{1}{2}[(-\frac{l}2, \frac{l}2),(-\frac{l+1}2,\frac{l+2}2)]_\alpha}}.
\end{align}
where $[(j_1,j_2),(r_1,r_2)]_\alpha$ denotes the number of multiple of $p$ among $\{ (q-1)j- (q+1)r + \alpha(m)\}_{j= j_1,j_1+1,\ldots ,j_2}^{r=r_1,r_1+1,\ldots, r_2}$. Up to sign the 1-loop determinant is simplified as
\begin{align}
&Z_{1-loop}[\sigma_0 , m]
\\
&= \prod_\alpha \big{(} [1+ i \alpha(\sigma_0)]^{[(- \frac{1}{2}, - \frac{1}{2}),(-\frac{1}2, \frac{1}2)]_\alpha} \prod_{l=0}^{\infty} [l+2+ i \alpha(\sigma_0)]^{A_\alpha} [l+2 - i \alpha(\sigma_0)]^{B_\alpha} \big{)} ,\nonumber
\\
&= \prod_{\alpha >0} \big{(} [1+ i \alpha(\sigma_0)]^{[(-\frac{1}{2}, \frac{1}{2}),(-\frac{1}{2}, \frac{1}{2})]_\alpha} \prod_{l=0}^\infty [l+2 + i \alpha(\sigma_0)]^{A_\alpha + B_{-\alpha}}[l+2 - i \alpha(\sigma_0)]^{A_{-\alpha}+B_\alpha}  \big{)},\nonumber
\\
&= \prod_{\alpha >0} \big{(} [1+ i \alpha(\sigma_0)]^{[(-\frac{1}{2}, \frac{1}{2}),(-\frac{1}{2}, \frac{1}{2})]_\alpha} \prod_{l=0;l+2 \pm \alpha(m)\in p \mathbb{Z}}^\infty [l+2 \pm i \alpha(\sigma_0)]  \prod_{l=0;q(l+2) \mp \alpha(m)\in p \mathbb{Z}}^\infty [l+2 \pm i \alpha(\sigma_0)]  \big{)},\nonumber
\\
&= \prod_{\alpha>0} \big{(} \prod_{n=1}^\infty (pn)^4 \big{)} (\frac{ p }{\pi})^2  \frac{ \sinh[\frac{\pi (\alpha(\sigma_0) + i \alpha(m))}p] \sinh[\frac{\pi (\alpha(\sigma_0) - i \alpha(m)^\sharp)}p]}{ ( \alpha(\sigma_0)^2)^{\delta_{\alpha(m),0}}}.
\end{align}
Here $A_\alpha,B_\alpha$ represent
\begin{align}
&A_\alpha : = [(-\frac{l+2}2, \frac{l}2),(-\frac{l+2}2, \frac{l+2}2)]_\alpha - \frac{1}{2} \big{(} [(-\frac{l+2}2,\frac{l+2}2),(-\frac{l}2,\frac{l}2)]_\alpha +[(-\frac{l}2,\frac{l}2),(-\frac{l+2}2 , \frac{l+2}2)]_\alpha \big{)} ,  \nonumber
\\
&B_\alpha : = [(-\frac{l+2}2, \frac{l}2),(-\frac{l}2, \frac{l}2)]_\alpha - \frac{1}{2} \big{(} [(-\frac{l+2}2,\frac{l+2}2),(-\frac{l}2,\frac{l}2)]_\alpha + [(-\frac{l}2,\frac{l}2),(-\frac{l+2}2 , \frac{l+2}2)]_\alpha \big{)} , \nonumber
\\
&A_\alpha +B_{-\alpha}  = [(-\frac{l+2}2, - \frac{l+2}2),(\frac{l+2}2, \frac{l+2}2)]_\alpha + [(-\frac{l+2}2, - \frac{l+2}2),(-\frac{l+2}2, -\frac{l+2}2)]_\alpha .
\end{align}
The condition $q(l+2)\pm \alpha(m)\in p \mathbb{Z}$ can be solved by $l+2 = p \mathbb{Z} \mp  \alpha(m)^\sharp$. An integer $0 \leq x^\sharp \leq p-1$ is uniquely  determined by the equation
\begin{align}
q x^\sharp - x \in p \mathbb{Z} .
\end{align}
Ignoring the overall infinity, the 1-loop determinant (including unfixed sign $\epsilon(m)$) becomes
\begin{align}
Z_{1-loop}[\sigma_0, m] = \epsilon(m)\prod_{\alpha>0}  \frac{ \sinh[\frac{\pi (\alpha(\sigma_0) + i \alpha(m))}p] \sinh[\frac{\pi (\alpha(\sigma_0) - i \alpha(m)^\sharp)}p]}{ ( \alpha(\sigma_0)^2)^{\delta_{\alpha(m),0}}}
\end{align}
For $G=U(N)$ roots $\alpha_{ij}$ are labeled by two integers $i\neq j$.
\begin{align}
\alpha_{ij}(\sigma_0) = \lambda_i - \lambda_j, \quad \sigma_0=\textrm{diagnal}(\lambda_1 \ldots \lambda_N).
\end{align}
We choose positive roots as $\alpha_{ij}$ with $i<j$. The partition function is given by $Z^{L(p,q)}(k):=\sum_m Z^{L(p,q)}(k,m)$ where
\begin{align}
&Z^{L(p,q)}(k,m) =  \int d \sigma_0 \exp(i S_0 [\sigma_0, m]) Z_{1-loop}[\sigma_0, m], \nonumber
\\
&=\epsilon(m) \int d^N  \lambda e^{- i \frac{\pi \hat{k} } p \sum_i (\lambda_i^2 - q^* m_i^2) } \prod_{\alpha>0}\sinh[\frac{\pi(\alpha \cdot \sigma_0 + i  \alpha \cdot m)}p ]\sinh[\frac{\pi ( \alpha \cdot \sigma_0 - (\alpha \cdot m)^\sharp )}p], \nonumber
\\
&=\epsilon(m) \epsilon'(m)\int d^N  \lambda e^{- i \frac{\pi \hat{k} } p \sum_i (\lambda_i^2 - q^* m_i^2) }  \prod_{\alpha>0}\sinh[\frac{\pi \alpha \cdot (\sigma_0 + i m)}p ]\sinh[\frac{\pi \alpha \cdot (\sigma_0 - i q^* m)}p]. \label{partiton on lens from localization}
\end{align}
We integrate $\sigma_0$ with Vandermonde measure on the broken symmetry $H(m) \in U(N)$ by the holonomy, $[H(m),m]=0$.
\begin{align}
d\sigma_0 = \int d^N  \lambda \prod_{i<j;m_i = m_j}(\lambda_i - \lambda_j)^2.
\end{align}
To understand this measure, consider  $N=2$ case for simplicity. If $m_1 = m_2$,  $\sigma_0$ can be an arbitrary $2 \times 2$ hermitian matrix and we must integrate over all of them. Since the integrand is invariant under the adjoint action of $U(2)$ on $\sigma_0$, we can replace the integral by an integral over a Cartan subalgebra $\sigma_0 = \textrm{diagonal}(\lambda_1, \lambda_2)$ with additional measure $(\lambda_1 -\lambda_2)^2$. The situation is different when $m_1 \neq m_2$. In that case $\sigma_0$ must be a Cartan to commute with $m$ and thus we integrate over a Cartan subalgebra without any additional measure.
\\
In the last line in \eqref{partiton on lens from localization}, we use the fact that $q^* x - x^\sharp \in p \mathbb{Z}$.
Although we can't prove it explicitly, we will assume that $\epsilon(m)\epsilon'(m)=1$. For $q= 1$ case, it can be easily shown that $\epsilon(m)=\epsilon(m')=1$ and thus our assumption is correct.
Under the assumption, we find exact match between \eqref{partiton on lens from localization} and the known results \eqref{known results 2-2} up to irrelevant overall contant taking account of  \eqref{redundance}. For Wilson loop along the loop $C_{p,q}$, we have
\begin{align}
\langle W_R(C_{p,q}) \rangle = &\frac{1}{Z^{L(p,q)}} \sum_m  \int d^N  \lambda e^{- i \frac{\pi \hat{k} } p \sum_i (\lambda_i^2 - q^* m_i^2) } \Tr_R \exp[\textrm{sign}(q)\frac{2 \pi ( q\sigma_0- i m)}p]  \nonumber
\\
&\times \prod_{\alpha>0}\sinh[\frac{\pi \alpha \cdot (\sigma_0 + i m)}p ]\sinh[\frac{\pi \alpha \cdot (\sigma_0 - i q^* m)}p].
\end{align}
For $q=-1$, it matches with known results \eqref{known results 2-3}.

\section{Discussion}
In this paper, we apply the localization technique used in \cite{Kapustin:2009kz} to Chern-Simons theory on more general manifolds than $S^3$. Our result is in a good agreement with  known results which are obtained in different ways.  There haver been several attempts to calculate  invariants of  Chern-Simons theory by path integral approach \cite{Blau:1993tv,Blau:2006gh,Beasley:2005vf,Beasley:2009mb}. Our localization method is relatively new path integral  approach to Chern-Simons theories and may give new insights on topology of three dimensional manifold. When calculating an one-loop determinant, we use the knowledge on spectrums of some differential operators. This may give some hints on the relation between spectrums of differential operators and topology of the manifold. In $S^2 \times S^1$ case, from the fact that the classical action $S_0$ for the saddle points with flat gauge connection is zero we can easily conclude that Wilson loop expectation value and partition function do not depends on $k$. The property is closely related to the fact that $\epsilon$ is not proportional to $\epsilon'$ in the Killing spinor equations, $\nabla_\mu \epsilon = \gamma_\mu \epsilon'$. From these observations, one may obtain some interesting mathematical results relating properties of Killing spinor on a manifold and topology of the space.
We consider $S^2 \times S^1$ and other lens spaces separately because our approach to the two cases are different. Our approach to $S^2 \times S^1$ may be generalized to $\Sigma_g \times S^1$ as in \cite{Blau:1993tv} where $\Sigma_g$ is a Riemann surface with genus $g$. On the other hand, our approach to the lens spaces with $p>0$ can be generalized to the more general spaces which can be obtained by discrete quotient of simpler space. Difficult thing in applying the localization method to more general manifolds is that we need many additional structures (metric and Killing spinor) to topological spaces. We are wondering weather there are any systematic way of  giving these structures to three manifolds from it surgery description.

\subsection*{Acknowledgments}
We would like to thank Takao Suyama for valuable discussions and comments on the manuscript. We also thank Seok Kim, Eunkyung Koh and Hee-Cheol Kim for useful discussions.

\appendix

\section{Spectrum of vector Laplcian and Dirac operator on $S^3$ and its $U(1)_Q$ charge}
$S^3$ can be considered as a group manifold $SU(2)$.
\begin{align}
\left(
            \begin{array}{cc}
              x & y \\
              -\bar{y} & \bar{x} \\
            \end{array}
          \right) \in SU(2), \quad |x|^2 + |y|^2 =1.
\end{align}
Functions on $S^3$ can be decomposed into irreducible representation of $SU(2)_L \times SU(2)_R$.
\begin{align}
\sum_{l\geq 0}[(l_L, l_R)= (\frac{l}2 ,\frac{l}2)]
\end{align}
Generators $(\vec{L}, \vec{R} )$ of $SU(2)_L \times SU(2)_R$ are given by
\begin{align}
&L_+ = x \partial_y - \bar{y}\partial_{\bar{x}}, \quad L_- = y \partial_x - \bar{x} \partial_{\bar{y}}, \nonumber
\\
&L_3 = \frac{1}{2} x \partial_x - \frac{1}{2} y\partial_y -\frac{1}{2} \bar{x}\partial_{\bar{x}} + \frac{1}{2}\bar{y}\partial_{\bar{y}}, \nonumber
\\
&R_+ = \bar{y}\partial_x - \bar{x} \partial_y, \quad R_- = - y \partial_{\bar{x}} + x\partial_{\bar{y}}, \nonumber
\\
&R_3 = - \frac{1}{2}x \partial_x - \frac{1}{2} y \partial_y + \frac{1}{2}\bar{x}\partial_{\bar{x}} + \frac{1}{2} \bar{y} \partial_{\bar{y}}. \label{generator of Su(2)*Su(2)}
\end{align}
Here $L_\pm := L_1 \pm i L_2$ and others are defined in a similar way. Divergenceless vector fields on $S^3$ is decomposed as
\begin{align}
\sum_{l \geq 0} [(j_L , l_R) = (\frac{l}2+ 1, \frac{l}2 )] \oplus [(j_L , l_R) = (\frac{l}2, \frac{l}2 +1)]
\end{align}
$j_L$ denotes the total angular momentum $\vec{J}_L = \vec{L}+\vec{S}_v$. Expanding a vector field $\vec{V}$ on $S^3$ in terms of left invariant vector fields $\vec{V}= V^i l_i (l_i  :=  2 i L_i )$, spin operator $\vec{S}_v$ acts on the vector indices $( S=1 )$, $(S^i_v \cdot \vec{V})^j := - i {\epsilon^{ij}}_k V^k$. (Hodge) Laplacian on divergenceless vector fields is given by
\begin{align}
\nabla^2 = 4 \vec{L}\cdot \vec{L} + 4 \vec{L}\cdot \vec{S}_v + 4.
\end{align}
Thus the spectrum of the operator is given by $\{(l+2)^2\}_{l=1,2,\ldots}$ with degeneracy $2(l+1)(l+3)$. Dirac operator in the left invariant frame $\{l_i\}$ is given by
\begin{align}
i \gamma^\mu \nabla_\mu = - 4 \vec{L}\cdot \vec{S}_{s} - \frac{3}2,
\end{align}
where the spin operator $\vec{S}_s$ acts on the spinor indices $(S=\frac{1}2)$, $S^i_{s} := \gamma^i /2$. Thus eigen spinor of dirac operator is decomposed as
\begin{align}
&\sum_{l\geq 0 }[(j_L, l_R) = (\frac{l+1}2, \frac{l}2)]\oplus [(j_L , l_R) = (\frac{l}2 ,\frac{l+1}2) ].
\end{align}
The eigenvalues are $-l- \frac{3}2$ and $l+\frac{3}2$ with degeneracy $(l+1)(l+2)$ for each. The Killing spinor considered in \cite{Kapustin:2009kz} corresponds to $|j_L = \frac{1}{2},l_R =0;(j_L)_z = \frac{1}{2}, (l_R)_z=0\rangle$.
\\
Now let's find the $U(1)_Q$ charge for each spectrum. Note that
\begin{align}
&Q[L_\pm]=Q[(S_v)_\pm] = Q[(S_s)_\pm] =\pm(q-1) , \quad Q(R_\pm) =\mp (q+1). \label{U(1) charge for rasing lowering}
\end{align}
The charge  of $L_\pm, R_\pm $ can be easily read from the explicit form \eqref{generator of Su(2)*Su(2)}. Using the fact $Q[| (s_v)_z = \pm 1\rangle  = \frac{1}{\sqrt{2}} (l_i \pm i l_2)] = \pm (q-1)$ and $Q[| (s_v)_z =  0\rangle = l_3] =0$, the charge of $(S_v)_\pm$ is determined. The connection between vector and bilinear of spinor determine the charge of $(S_s)_\pm $. From \eqref{U(1) charge for rasing lowering}  the charge for divergenceless vector fields in $[(j_L, l_R)]$ is given by
\begin{align}
Q = (q-1)(j_L)_z - (q+1)(l_R)_z
\end{align}
Similarly, the charge for the spinor in $[j_L, l_R]$ is
\begin{align}
Q= (q-1)\big{(}(j_L)_z -\frac{1}2 \big{)} - (q+1)(l_R)_z .
\end{align}
We assign the $U(1)$ charge zero to $|(s_s)_z =\frac{1}{2}\rangle $ so that the Killing spinor is invariant under the $\mathbb{Z}_p\in U(1)_Q$.

\section{Monopole Harmonics on $S^2$ }
In this section, we summarize some relevant facts on monopole (scalar, vector, spinor) harmonics. See \cite{Kim:2009wb,Benna:2009xd} for more details.
Scalar monopole harmonics $Y_{qjm}$, ($j= |q| , |q|+1, \ldots$ and $m=-j,-j+1,\ldots, j-1 ,j$) satisfies
\begin{align}
\bar{D}_i \bar{D}^i Y_{qjm} = -(j(j+1)-q^2) Y_{qjm}. \label{monopole harmonics 1}
\end{align}
Here $\bar{D}_i$ is the covariant derivative on $S^2$ in monopole background with charge $q$. From dirac quantization condition, charge $q$ takes a half-integer value. Vector monopole harmonics $\vec{C}^{\pm}_{qjm}$ on $S^2$  are
\begin{align}
&\vec{C}^{+}_{qjm}, \quad j = |q|-1(\textrm{for $q\geq 1$}),|q|, \ldots,
\\
&\vec{C}^{-}_{qjm}, \quad j=|q|+1,|q|+2,\ldots,
\end{align}
These satisfy
\begin{align}
&\vec{\bar{D}}\cdot \vec{C}^{\pm}_{qjm} = s_{\pm}Y_{qjm} (j\geq |q|+1), \quad \vec{\bar{D}}\cdot \vec{C}^{+}_{qjm} = s_{+}Y_{qjm} (j=|q|),  \quad \vec{\bar{D}}\cdot \vec{C}^{+}_{qjm} =0 (j=|q|-1), \nonumber
\\
&\bar{D}_1 (C^{\pm}_{qjm})_2 - \bar{D}_2 (C^{\pm}_{qjm})_1 = \mp s_{\pm} Y_{qjm}(j\geq |q|+1),   \nonumber
\\
&\bar{D}_1 (C^{+}_{qjm})_2 - \bar{D}_2 (C^{+}_{qjm})_1 = - s_{+} Y_{qjm}(j= |q|-1,|q|).
\end{align}
Here $1,2$ denote vielbein indices of the $S^2$. $s_\pm$ is defined by
\begin{align}
s_\pm := \sqrt{\frac{j(j+1)-q^2 \pm |q|}2}.
\end{align}
Vector fields $\vec{V}_{qjm}$ satisfying $\vec{\bar{D}} \cdot \vec{V}_{qjm}=0$ are
\begin{align}
&\vec{V}_{qjm} = a_+  \vec{C}^+_{qjm} + a_-  \vec{C}^-_{qjm}, \quad j\geq |q|+1, \nonumber
\\
&\vec{V}_{qjm} = \vec{C}^+_{qjm}, \quad j=|q|-1.
\end{align}
Here $a_\pm$ are uniquely determined by the equations (with the condition $a_+ \geq 0 $)
\begin{align}
a_+ s_+ + a_- s_- =0, \quad  (a_+)^2 + (a_-)^2=1.
\end{align}
They satisfy
\begin{align}
&\bar{D}_1 (V_{qjm})_2 - \bar{D}_2 (V_{qjm})_1  = \kappa_{qj} Y_{qjm}, \quad j \geq |q|+1, \nonumber
\\
&\bar{D}_1 (V_{qjm})_2 - \bar{D}_2 (V_{qjm})_1 =0, \quad j=|q|-1. \label{monopole harmonics 2}
\end{align}
Here $\kappa_{qj}$ is
\begin{align}
\kappa_{qj} : = \frac{\sqrt{(j(j+1)-q^2)^2 - q^2}}{\sqrt{j(j+1)-q^2}}.
\end{align}
Eigenspinor $\Psi^\pm_{qjm}, \Psi^0_{q j m }$ of the Dirac operator on a unit two sphere in the monopole background is given by ($i=1,2$ is the vielbein indices for $S^2$)
\begin{align}
&i \gamma^i \bar{D}_i \Psi^{\pm}_{qjm} = \pm \mu_{jq} \Psi^{\pm}_{qjm}, \quad j = |q|+\frac{1}{2},|q|+\frac{3}2 \ldots, \nonumber
\\
&i \gamma^i \bar{D}_i \Psi^{0}_{qjm} =0, \quad j=  |q|-\frac{1}{2}, \label{monopole harmonics 3}
\end{align}
with $\mu_{jq}:= \frac{1}2 \sqrt{(2j+1)^2 - 4 q^2} $.
They satisfy
\begin{align}
&\gamma^3 \Psi^{\pm}_{qjm} = \Psi^{\mp}_{qjm}, \nonumber
\\
&\gamma^3 \Psi^{0}_{qjm} =\textrm{sign}(q) \Psi^{0}_{qjm}. \label{monopole harmonics 4}
\end{align}
Here $\gamma^3$ is given by $\gamma^1 \gamma^2 = i \gamma^3$.
\providecommand{\href}[2]{#2}\begingroup\raggedright\endgroup

\begin{thebibliography}{10}

\bibitem{Pestun:2007rz}
  V.~Pestun,
  ``Localization of gauge theory on a four-sphere and supersymmetric Wilson
  loops,''
  arXiv:0712.2824 [hep-th].

\bibitem{Erickson:2000af}
  J.~K.~Erickson, G.~W.~Semenoff and K.~Zarembo,
  ``Wilson loops in N = 4 supersymmetric Yang-Mills theory,''
  Nucl.\ Phys.\  B {\bf 582}, 155 (2000)
  [arXiv:hep-th/0003055].

\bibitem{Drukker:2000rr}
  N.~Drukker and D.~J.~Gross,
  ``An exact prediction of N = 4 SUSYM theory for string theory,''
  J.\ Math.\ Phys.\  {\bf 42}, 2896 (2001)
  [arXiv:hep-th/0010274].


\bibitem{Kapustin:2009kz}
  A.~Kapustin, B.~Willett and I.~Yaakov,
  ``Exact Results for Wilson Loops in Superconformal Chern-Simons Theories with
  Matter,''
  arXiv:0909.4559 [hep-th].

\bibitem{Marino:2002fk}
  M.~Marino,
  ``Chern-Simons theory, matrix integralsf, and perturbative three-manifold
  invariants,''
  Commun.\ Math.\ Phys.\  {\bf 253}, 25 (2004)
  [arXiv:hep-th/0207096].


\bibitem{Aganagic:2002wv}
  M.~Aganagic, A.~Klemm, M.~Marino and C.~Vafa,
  ``Matrix model as a mirror of Chern-Simons theory,''
  JHEP {\bf 0402}, 010 (2004)
  [arXiv:hep-th/0211098].

\bibitem{Dolivet:2006ii}
  Y.~Dolivet and M.~Tierz,
  ``Chern-Simons matrix models and Stieltjes-Wigert polynomials,''
  J.\ Math.\ Phys.\  {\bf 48}, 023507 (2007)
  [arXiv:hep-th/0609167].



\bibitem{brini:2008}
  A.~Brini, L.~Griguolo, D.~Seminara and A.~Tanzini,
  ``Chern-Simons theory on L(p,q) lens spaces and Gopakumar-Vafa duality,''
  arXiv:0809.1610 [math-ph].

\bibitem{Griguolo:2006kp}
  L.~Griguolo, D.~Seminara, R.~J.~Szabo and A.~Tanzini,
  ``Black holes, instanton counting on toric singularities and q-deformed
  two-dimensional Yang-Mills theory,''
  Nucl.\ Phys.\  B {\bf 772}, 1 (2007)
  [arXiv:hep-th/0610155].



\bibitem{Witten:1988hf}
  E.~Witten,
  ``Quantum field theory and the Jones polynomial,''
  Commun.\ Math.\ Phys.\  {\bf 121}, 351 (1989).

\bibitem{hansen-2004-5}
  S.~K. Hansen and T. Takata,
  ``Reshetikhin-Turaev invariants of Seifert 3-manifolds for classical   simple Lie algebras, and their asymptotic expansions,''
  J.Knot Theory Ramifications {\bf 13}, 617 (2004) [arXiv:math.GT/0209403].

\bibitem{Halmagyi:2007rw}
  N.~Halmagyi and T.~Okuda,
  ``Bubbling Calabi-Yau geometry from matrix models,''
  JHEP {\bf 0803}, 028 (2008)
  [arXiv:0711.1870 [hep-th]].



\bibitem{Schwarz:2004yj}
  J.~H.~Schwarz,
  ``Superconformal Chern-Simons theories,''
  JHEP {\bf 0411}, 078 (2004)
  [arXiv:hep-th/0411077].

\bibitem{Kim:2009wb}
  S.~Kim,
  ``The complete superconformal index for N=6 Chern-Simons theory,''
  Nucl.\ Phys.\  B {\bf 821}, 241 (2009)
  [arXiv:0903.4172 [hep-th]].

\bibitem{Hosotani:1988bm}
  Y.~Hosotani,
  Annals Phys.\  {\bf 190}, 233 (1989).



\bibitem{Blau:1993tv}
  M.~Blau and G.~Thompson,
  ``Derivation of the Verlinde formula from Chern-Simons theory and the G/G
  model,''
  Nucl.\ Phys.\  B {\bf 408}, 345 (1993)
  [arXiv:hep-th/9305010].

\bibitem{Blau:2006gh}
  M.~Blau and G.~Thompson,
  ``Chern-Simons theory on S**1-bundles: Abelianisation and q-deformed
  Yang-Mills theory,''
  JHEP {\bf 0605}, 003 (2006)
  [arXiv:hep-th/0601068].

\bibitem{Beasley:2005vf}
  C.~Beasley and E.~Witten,
  ``Non-abelian localization for Chern-Simons theory,''
  J.\ Diff.\ Geom.\  {\bf 70}, 183 (2005)
  [arXiv:hep-th/0503126].

\bibitem{Beasley:2009mb}
  C.~Beasley,
  ``Localization for Wilson Loops in Chern-Simons Theory,''
  arXiv:0911.2687 [hep-th].

\bibitem{Benna:2009xd}
  M.~K.~Benna, I.~R.~Klebanov and T.~Klose,
  ``Charges of Monopole Operators in Chern-Simons Yang-Mills Theory,''
  arXiv:0906.3008 [hep-th].



\end{thebibliography}
\end{document}